\documentclass[letterpaper,showpacs,preprintnumbers,amsmath,amssymb,nofootinbib,notitlepage]{revtex4-1}
\usepackage{graphicx}
\usepackage[usenames,dvipsnames]{color}
\usepackage[colorlinks,citecolor=blue]{hyperref}
\usepackage{amsmath}
\usepackage{bbm}
\usepackage{amsfonts}
\usepackage{amssymb}
\usepackage{latexsym}
\usepackage{graphicx}
\usepackage[english]{babel}
\usepackage{multirow}
\usepackage{float}
\usepackage{url}
\usepackage{slashed}
\usepackage{xcolor}
\usepackage{enumerate}
\usepackage{makecell}
\usepackage{diagbox}
\usepackage{makecell}

\newcommand{\GeV}{\,{\rm GeV}}
\newcommand{\MeV}{\,{\rm MeV}}

\newcommand{\be}{\begin{equation}}
\newcommand{\ee}{\end{equation}}
\newcommand{\ba}{\begin{array}}
\newcommand{\ea}{\end{array}}
\newcommand{\bea}{\begin{eqnarray}}
\newcommand{\eea}{\end{eqnarray}}

\usepackage{ulem,fancyvrb}
\usepackage{xcolor}

\begin{document}
\title{Probing invisible decay of dark photon at BESIII and future STCF via monophoton searches}

\author{Yu Zhang$^{1,2}$~}  \email{dayu@nju.edu.cn, dayu@ahu.edu.cn}

\author{Wei-Tao Zhang$^{1}$~}

\author{Mao Song$^2$~} 

\author{Xue-An Pan$^2$~} 

\author{Zhong-Ming Niu$^{1,2}$~} 

\author{Gang Li$^2$~}

\affiliation{
$^1$ Institutes of Physical Science and Information Technology,
Anhui University, Hefei 230601, China \\ 
$^2$ School of Physics and Materials Science, Anhui University, Hefei 230039,China }

\begin{abstract}
We propose to search the monophoton events  at the BESIII detector 
and future Super Tau Charm Factory to probe 
the sub-GeV dark photon decay into lighter dark matter.
We compute the cross section due to the dark photon associated a standard model photon
production, and study the corresponding standard model irreducible/reducible
backgrounds. By using the data  about 17 fb$^{-1}$ collected at the BESIII detector since 2011,
we derive new leading limits of the mixing strength $\varepsilon$,
$\varepsilon\lesssim(1.1-1.6)\times 10^{-4}$,
in the mass range of 0.04 GeV $\lesssim m_{A^\prime} \lesssim$ 3 GeV.
With 30 ab$^{-1}$ data, STCF running at $\sqrt{s} = 2$ GeV,
can probe $\varepsilon$ down to 5.1$\times 10^{-6}$ when $m_{A^\prime}=1$ GeV.
For models of scalar and fermionic light thermal dark matter 
production via dark photon, we present the  constrains
on the dimensionless dark matter parameter 
$y=\varepsilon^2\alpha_D(m_\chi/m_{A^\prime})^4$ as function of the DM mass $m_{\chi}$ at BESIII and future STCF,
conventionally assuming the dark coupling constant $\alpha_D=0.5$ and $m_{A^\prime}=3 m_{\chi}$.
We find that BESIII can exclude model of scalar, Majorana, and pseudo-Dirac (with a small splitting) DM for the 
mass region 0.03$\sim$1 GeV, 0.04$\sim$1 GeV and 0.4$\sim$1 GeV respectively.  
For values $\alpha_D\lesssim 0.005$, combining the results from 2 GeV STCF with 30 ab$^{-1}$
data and BaBar, one can exclude the above three DM models in the mass region 0.001 GeV $\lesssim m_{\chi}
\lesssim$ 1 GeV.

\end{abstract}

\maketitle

\section{Introduction}
\label{sec:intro}
To investigate the nature of dark matter (DM) particle  is one of 
the most pressing issues in modern physics.
So far, we have only been able to probe the DM through its
gravitational effects with visible matter.
It is widely postulated that DM interact very weakly
with ordinary matter, since terrestrial searches haven't yielded any results yet.
An exciting attempt is that, besides the gravity, one can introduce an extra $U_D(1)$ force carrier,
also referred to as dark photon $A^\prime$, providing a natural scenario for DM interactions,
which is neutral under the SM gauge symmetries, but couples
to the SM photon via kinetic mixing \cite{Holdom:1985ag}.
The kinetic mixing term can be described as
\be
{\cal L}_{\rm kinetic\ mixing}=-\frac{\varepsilon}{2}F^\prime_{\mu\nu} F^{\mu\nu}  
\ee
and parametrized by the mixing strength $\varepsilon \ll 1$,
where $F^\prime_{\mu\nu}=\partial_\mu A^\prime_{\nu} -
\partial_\nu A^{\prime}_{\mu}$ is the field strength of $A^\prime$ ,
resulting in the interaction 
\be
{\cal L}_{\rm int}=\varepsilon e A^\prime_\mu J^\mu_{em},  
\ee
of dark photon with the  electromagnetic current $J^\mu_{em}$ with 
a strength $\varepsilon e$, where $e$ is 
the electromagnetic coupling.  
In order to explain observational
astroparticle anomalies, the dark photon should be relatively light,
with a mass in the MeV to GeV range \cite{ArkaniHamed:2008qn}.
Futhermore, a sub-GeV $A^\prime$ with $\varepsilon\simeq 10^{-3}$ can also 
explain the 3.6$\sigma$ deviation from the SM prediction of the muon anomalous magnetic
moment $(g-2)_\mu$ \cite{Bennett:2006fi,Pospelov:2008zw}. 

The decay modes of the dark photon depend on its mass and couplings,
as well as on the particle spectrum of the dark sector.
Since there are no firm predictions for the dark photon,
various experiments have been searched for it over a wide 
range of its mass and decay modes.
If the dark photon is lightest in the dark sector,
its dominant decays are to the visible SM particles.
The searches for such dark photons with the mass below a few GeV
have been mainly performed in beam dump \cite{Riordan:1987aw, Bross:1989mp, Andreas:2012mt} , fixed target \cite{Merkel:2014avp, Abrahamyan:2011gv}, 
collider \cite{Babusci:2012cr, Babusci:2014sta, Agakishiev:2013fwl, Adare:2014mgk, Lees:2014xha, Ablikim:2017aab}
 and 
rare meson decay \cite{Batley:2015lha, Adlarson:2013eza} experiments using narrow peak in the $e^+e^-$ or $\mu^+\mu^-$ invariant mass spectra.
These limits are obtained assuming that the dark photon
dominantly decays to the visible SM particles, and
will be invalid if there are low-mass invisible degrees of freedom.
If the lowest-mass  DM states $\chi$ is sufficiently light,
in particular $m_\chi<m_{A^\prime}/2$, the $A^\prime$ 
would predominantly decay invisibly into the DM particles
provided that coupling  $e_D > \varepsilon e$.
$e_D$ is the coupling constant of the $U_D(1)$ gauge
interactions. There are limits on 
invisible decays of the dark photon from 
kaon decays by the E787 \cite{Adler:2001xv} and E949 \cite{Artamonov:2009sz} experiments,
$\pi^0$ decays by NA62 \cite{CortinaGil:2019nuo} experiment,
searches for missing energy events in electron-nucleus scattering by 
NA64 \cite{Banerjee:2016tad, Banerjee:2017hhz, NA64:2019imj} experiment, 
and monophoton searches by BaBar \cite{Lees:2017lec}.

In this paper, we focus on the search for the invisible decay of dark photon 
at the BESIII detector and future Super Tau Charm Factory (STCF).
The BESIII detector is operated at the Beijing Electron Positron Collider (BEPCII),
which is a double ring $e^+e^-$ collider running at the center-of-mass (c.m.) energy
$\sqrt s$ from 2.0 to 4.6 GeV with a peaking luminosity of $10^{33}$ cm$^{-2}$s$^{-1}$.
The STCF is a projected electron-positron collider operating in the range of
center-of-mass energies from 2.0 to 7.0 GeV with the peak luminosity of 
about $10^{35}$ cm$^{-2}$s$^{-1}$ \cite{stcf1,Bondar:2013cja}.
We assume that the decay width of the $A^\prime$
is negligible compared to the experimental resolution, and that 
the invisible branching ratio Br$( A^\prime\to \chi\bar\chi)\simeq 100\%$.
The cleanest collider signature of such particles is the process
 $e^+e^-\to \gamma A^\prime$, followed by invisible decay of the $A^\prime$,
which is monochromatic single photon production accompanied by 
significant missing energy and momentum. 
The monophoton signal has been investigated previously at BESIII detector\cite{Zhu:2007zt,Yin:2009mc,Liu:2018jdi}. 
Here we use the monophoton signature to probe invisible decay of dark photon at the BESIII detector and STCF.

The rest of the paper is organized as follows. 
In  Sec. \ref{sec:sb},   we study the monophoton
signature arising from dark photon production  and from the SM backgrounds.
The  results on the searches for invisible decay of dark photon 
at BESIII and future STCF are presented in Sec. \ref{sec:dplimit}.
The constrains on light thermal dark matter
are reported in Sec. \ref{sec:dmlimit}.
We summarize our findings in Sec. \ref{sec:sum}.

\section{Signals and Backgrounds}
\label{sec:sb}

\begin{figure}[htbp]
	\begin{centering}
		\includegraphics[width=0.4\columnwidth]{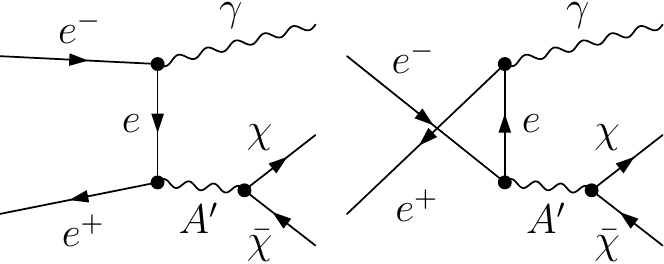}
		\caption{The Feynman diagrams for the production of an on-shell $A^\prime$ and a photon, in which we assume the
			$A^\prime$ subsequently decays to lighter dark matter.
		}
		\label{fig:eeaA} 
	\end{centering}
\end{figure}

At the electron colliders, the dark photon can be searched in the process
$e^+e^-\to \gamma A^\prime$, whose diagrams are shown in Fig.\ref{fig:eeaA}, with its subsequent decay
to lighter DM.

The differential cross section for an on-shell $A^\prime$ and a photon production 
process $e^+e^-\to \gamma A^\prime$  is 
\cite{Essig:2009nc}
\be
\frac{d\sigma_{\gamma A^\prime}}{dz_\gamma} = 
\frac{2\pi \varepsilon^2\alpha^2}{s}\left(1-\frac{m_{A^\prime}^2}{s}\right)
\frac{1+z_\gamma^2+\frac{4sm_{A^\prime}^2}{(s-m_{A^\prime}^2)^2}}{(1+z_\gamma)(1-z_\gamma)},
\ee
where $\alpha$ is the fine structure constant, $z_\gamma\equiv\cos\theta_\gamma$ with $\theta_\gamma$ 
being the relative angle between the electron beam axis and the photon momentum,
$s$ is the square of the center-of-mass energy, $m_{A^\prime}$ is the mass of the dark photon.
The photon energy $E_\gamma$ in the center-of-mass frame is related to the dark photon mass
as 
\be
E_\gamma = \frac{s-m_{A^\prime}^2}{2\sqrt{s}}.
\label{eq:egma}
\ee
The cross section after integrating the polar angle $\theta_\gamma$ 
is given as \cite{Essig:2009nc}
\bea
\label{eq:xs-signal}
\sigma_{\gamma A^\prime}=\frac{2\pi \varepsilon^2\alpha^2}{s}\left(1-\frac{m_{A^\prime}^2}{s}\right)
\left[\left(1+\frac{2sm_{A^\prime}^2}{(s-m_{A^\prime}^2)^2}\right){\cal Z}
-z_\gamma^{\rm max}+z_\gamma^{\rm min}\right],
\eea
where
\be
{\cal Z}=\ln\frac{(1+z_\gamma^{\rm max})(1-z_\gamma^{\rm min})}{(1-z_\gamma^{\rm max})(1+z_\gamma^{\rm min})} .
\ee

\begin{figure}[htbp]
	\begin{centering}
		\includegraphics[width=0.6\columnwidth]{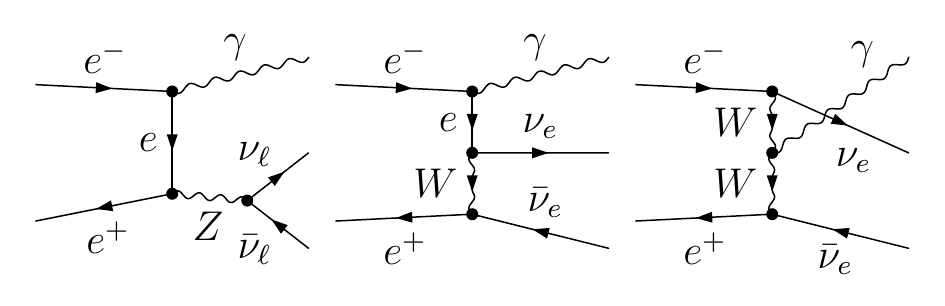}
		\caption{The Feynman diagrams for the SM processes $e^+ e^- \to \nu_\ell \bar \nu_\ell \gamma$,
			where $\nu_\ell = \nu_e, \nu_\mu, \nu_\tau$ are the three standard model neutrinos.
		}
		\label{fig:eevva} 
	\end{centering}
\end{figure}

The irreducible SM backgrounds to the monophoton signature at electron-positron
colliders are the $e^+ e^- \to \nu_\ell \bar \nu_\ell \gamma$ processes, 
where $\nu_\ell = \nu_e, \nu_\mu, \nu_\tau$ are the three standard model neutrinos. 
The corresponding Feynman diagrams are displayed in Fig.\ \ref{fig:eevva}. 
For electron neutrinos, both  $Z$-boson and  $W$-boson diagrams contribute; 
for the muon and tau neutrinos  only   $Z$-boson 
diagrams. For the electron-positron 
colliders running with GeV beam energy, the 
diagram mediated by two $W$ bosons can be safely eliminated in our analysis,
since it is suppressed by an additional $W$-boson propagator in comparison with
the other single $W$ or $Z$ mediator diagrams. 
The differential production cross section 
for the $e^+ e^- \to \nu \bar\nu \gamma$ processes 
mediated by a single $W/Z$ boson is given by 
\cite{Ma:1978zm} \cite{Gaemers:1978fe}
\be
{d\sigma \over d E_\gamma d z_\gamma}
=  {\alpha G_F^2  s_\gamma^2  
	\over 4 \pi^2  s E_\gamma (1 - z_\gamma^2) } f(\sin\theta_W)
\Bigg[ 1 + {E_\gamma^2 \over s_\gamma} (1 + z^2_\gamma)
\Bigg], 
\label{eq:irbg}
\ee
where $G_F$ is the Fermi constant, 
$f(\sin\theta_W) = 8 \sin^4\theta_W- 4 \sin^2\theta_W/3+1$ with 
$\theta_W$ being the weak  mixing angle. 
Here we have integrated over the momenta of the 
final state neutrinos and summed all three 
neutrino flavors. 

\begin{figure}[htbp]
	\begin{centering}
		\includegraphics[width=0.45\columnwidth]{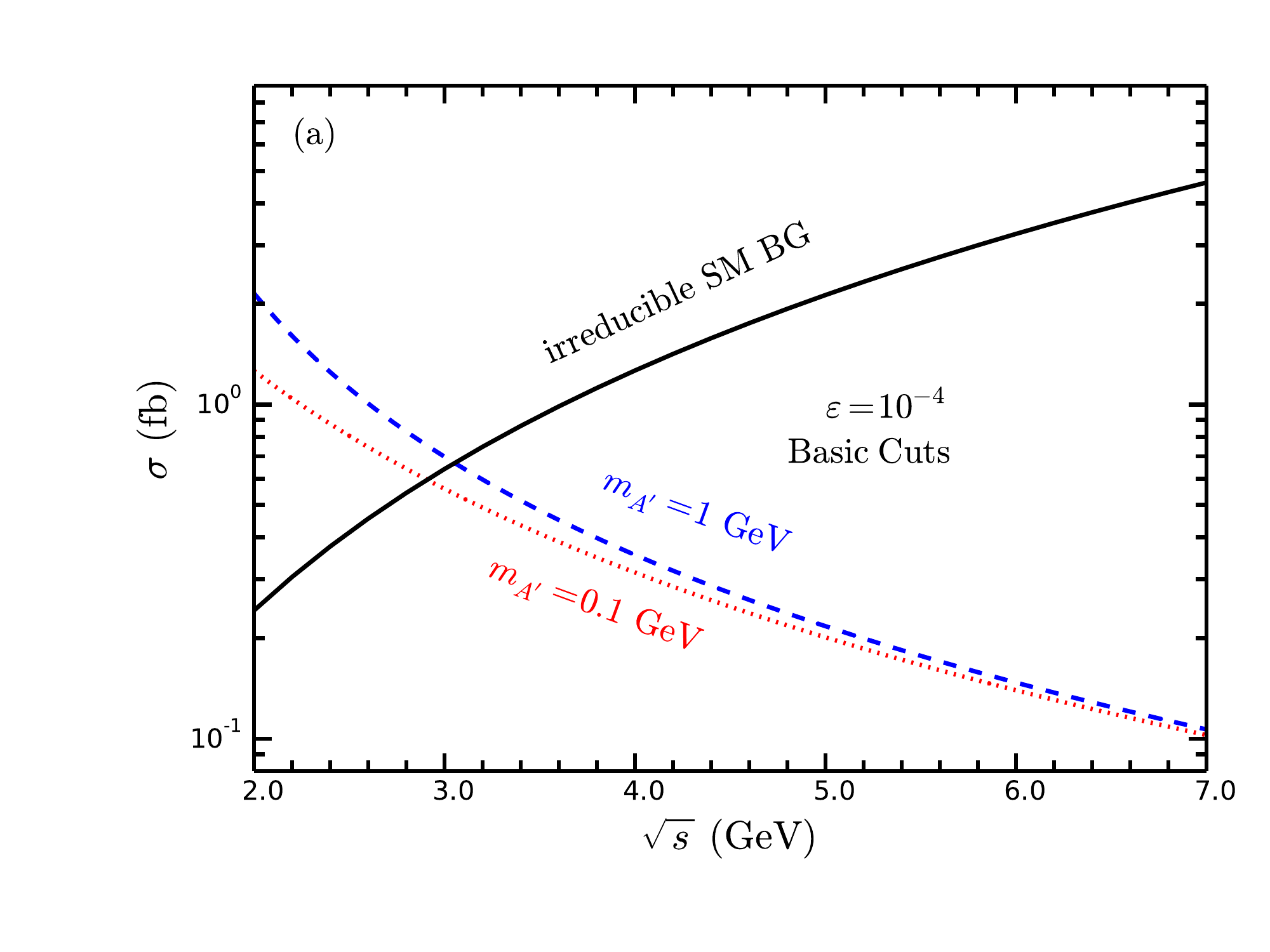}
		\includegraphics[width=0.45\columnwidth]{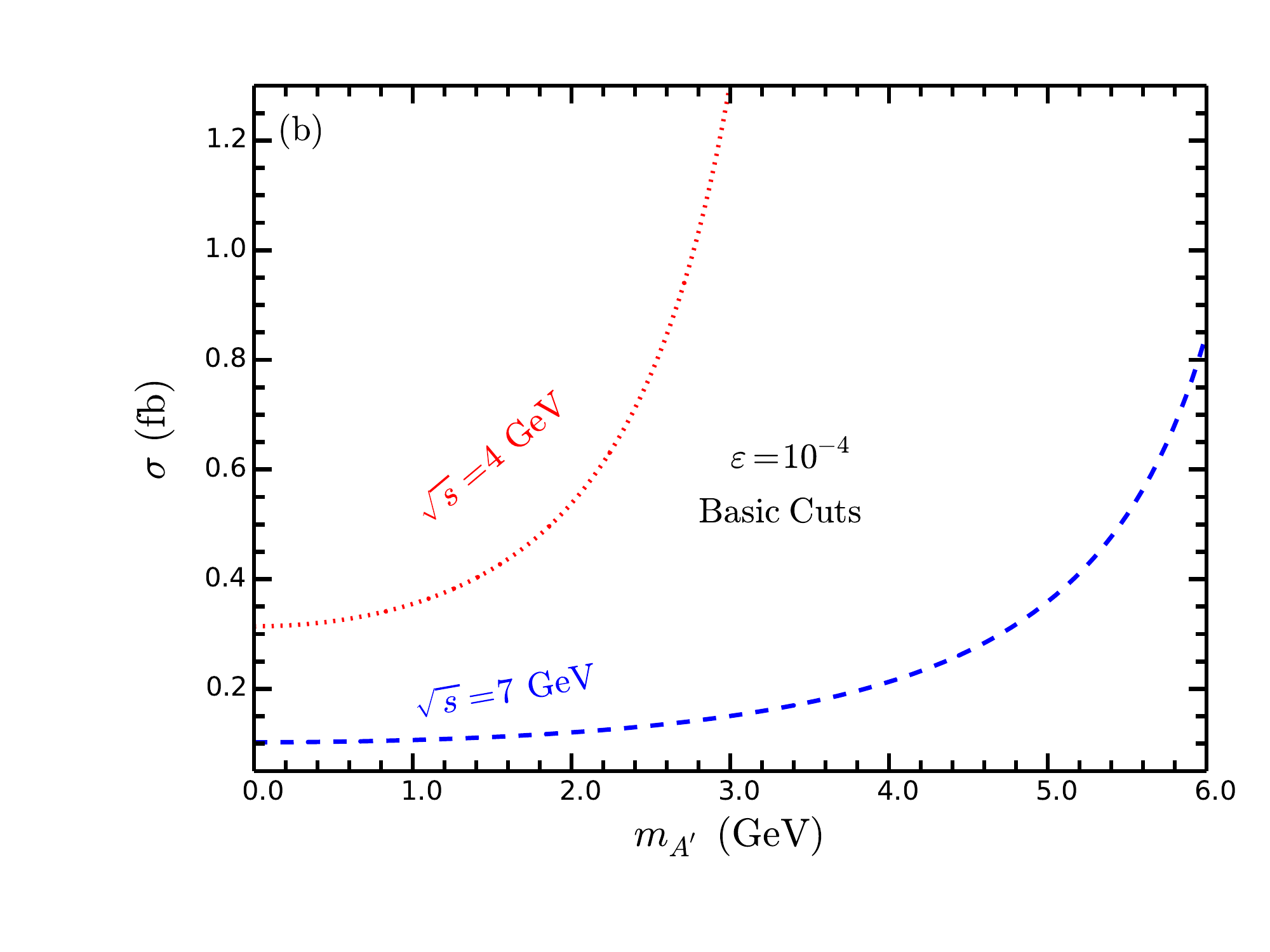}
		\caption{(a) The total cross section as a function of the collider energy for the irreducible SM background 
			$e^+ e^- \to \nu \bar\nu \gamma$, and for the production of an on-shell dark photon and a photon  with 
			$m_{A^\prime}$ = 0.1 GeV and $m_{A^\prime}$ = 1 GeV, respectively. (b) The total cross section as a function of the mass of  the dark photon for the production of an on-shell dark photon and a photon  with 
			$\sqrt s $ = 4 GeV and 	$\sqrt s $ = 7 GeV, respectively. All the results are obtained by adopting the ``basic cuts" and setting $\varepsilon=10^{-4}$.
		}
		\label{fig:xs-basic-cuts} 
	\end{centering}
\end{figure}

In Fig. \ref{fig:xs-basic-cuts} (a), we present the total cross section as a function of the collider energy
for the irreducible SM background, and for the production of an on-shell dark photon and a photon  with 
$m_{A^\prime}$ = 0.1 GeV and $m_{A^\prime}$ = 1 GeV, respectively.
For the final photon, we adopt the cuts:
$E_\gamma > 25 \MeV$ in the barrel ($|z_\gamma|<0.8$)  or 
$E_\gamma > 50 \MeV$ in the end-caps ($0.86<|z_\gamma|<0.92$),
following the cuts used by the BESIII Collaboration \cite{Ablikim:2017ixv},
which are defined as the ``basic cuts" hereafter. 
We can see that the production rates for dark photon associated with one SM photon 
drop rapidly when the colliding energy increases; 
however, the monophoton cross section due to the SM irreducible processes grows with the colliding energy.
Thus, electron collider with smaller colliding energy has a better sensitivity to 
search the invisible decay of dark photon when kinematics is accessible.
In Fig. \ref{fig:xs-basic-cuts} (b), we also provide the dependence of the total cross section for dark photon production 
on its mass when $\sqrt s$ = 4 GeV and 7 GeV with the basic cuts.
The production rates keep growing with the increment of the mass of dark photon,
which can be seen from Eq. (\ref{eq:xs-signal}) that the cross section is divergent when 
$\sqrt{s}\to m_{A^\prime}$ ($E_\gamma\to  0$).

Due to the limited detection capability of the subdetectors,
the  reducible backgrounds become important and should be 
investigated carefully.   
The reducible SM backgrounds mainly come from the $e^+e^-\to\gamma+{\slashed X}$ 
processes, where in the final state only one photon 
can be detected in the detectors, and ${\slashed X}$ denotes that the other 
particles are undetected because of the limitations of the detectors.
The dominate reducible backgrounds include the processes 
$e^+e^-\to f\bar{f}\gamma$ and $e^+e^-\to\gamma\gamma\gamma$
\footnote{The reducible background from the $e^+e^-\to\gamma\gamma$ 
process vanishes because the BESIII and STCF detectors
are arranged in a symmetric manner.},
which can be quite large with the final $f\bar{f}$ and $\gamma\gamma$
emitting in the solid angle region that is uncovered by detectors.
Especially, for the radiative Bhabha scattering process 
$e^+e^-\to e^+e^-\gamma$, when both final state electron 
and positron go along the beam directions,
the collinear singularity will arise in the $t$ channel diagrams,
and cause large cross section \cite{Mana:1986je, Actis:2009uq, Liu:2018jdi}.

Due to momentum conservation in the transverse direction and energy conservation,
the monophoton reducible background at the electron-positron
colliders can be removed by applying the detector cut \cite{Liu:2019ogn}:
\be
E_\gamma >E_b(\theta_\gamma)= \frac{\sqrt{s}}{(1+{\sin\theta_\gamma}/{\sin\theta_b})},
\label{eq:adv-cuts}
\ee
on the final state photon,
where the energy cut $E_b$ is the function of the polar angle $\theta_b$, and   $\theta_b$ denotes
the angle at the boundary of the sub-detectors.
We will collectively refer to the ``basic cuts" and cut (\ref{eq:adv-cuts}) as the ``advanced cuts" hereafter.
At the BESIII, 
we follow Ref. \cite{Liu:2018jdi},
and define the polar angel $|\cos\theta_b|=0.95$ 
after considering all the boundary of the subdetectors.
When $\theta_\gamma=\pi/2$, 
the energy cut $E_b$ achieves its minimum value 
$E_b^{\rm min} \simeq 0.24 \sqrt{s}$. 
In order to propose the sensitivity of STCF to dark photon, 
we assume that the sub-detectors of STCF have the same acceptance with the BESIII.

\begin{figure}[htbp]
	\begin{centering}
		\includegraphics[width=0.45\columnwidth]{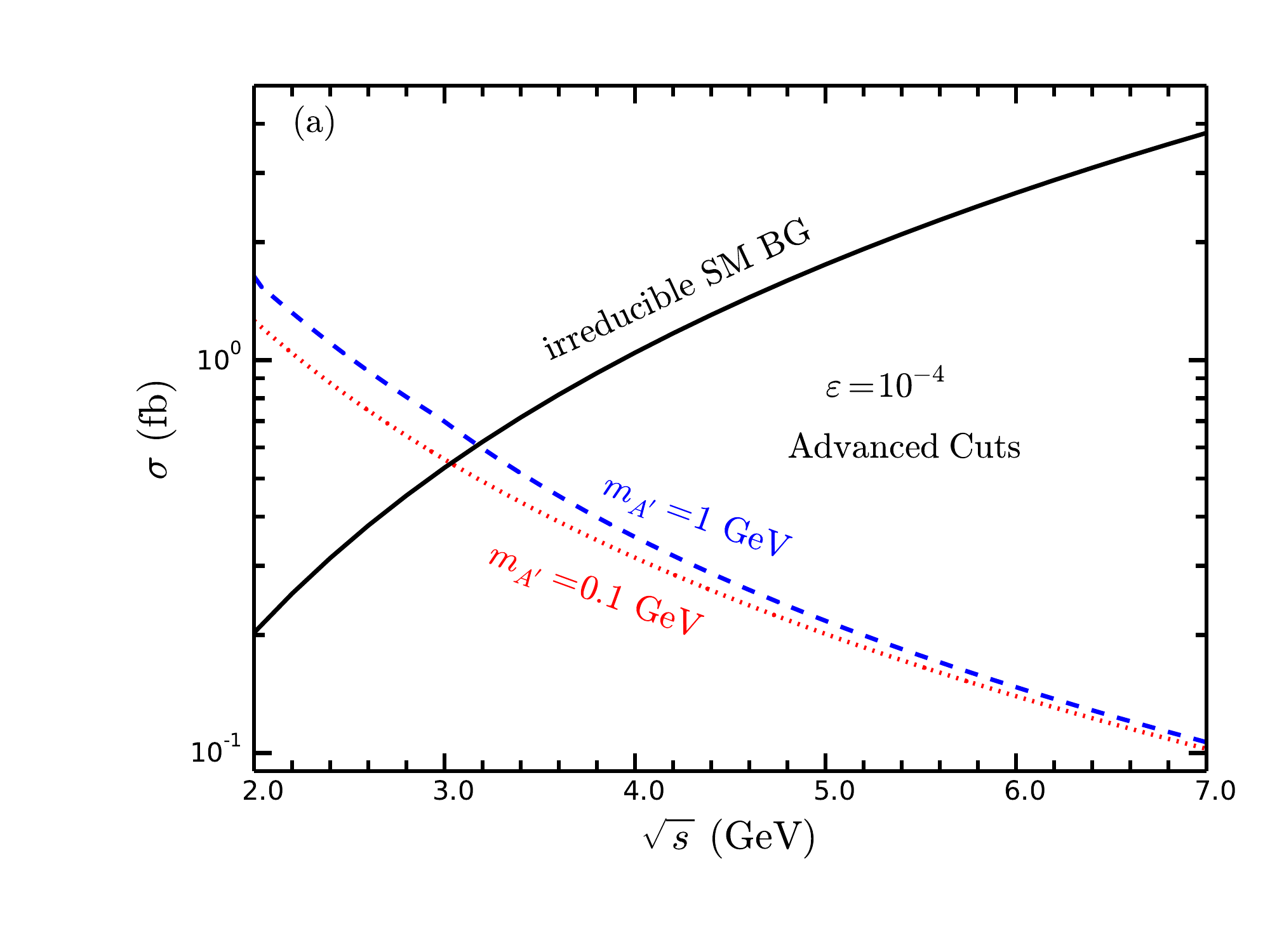}
		\includegraphics[width=0.45\columnwidth]{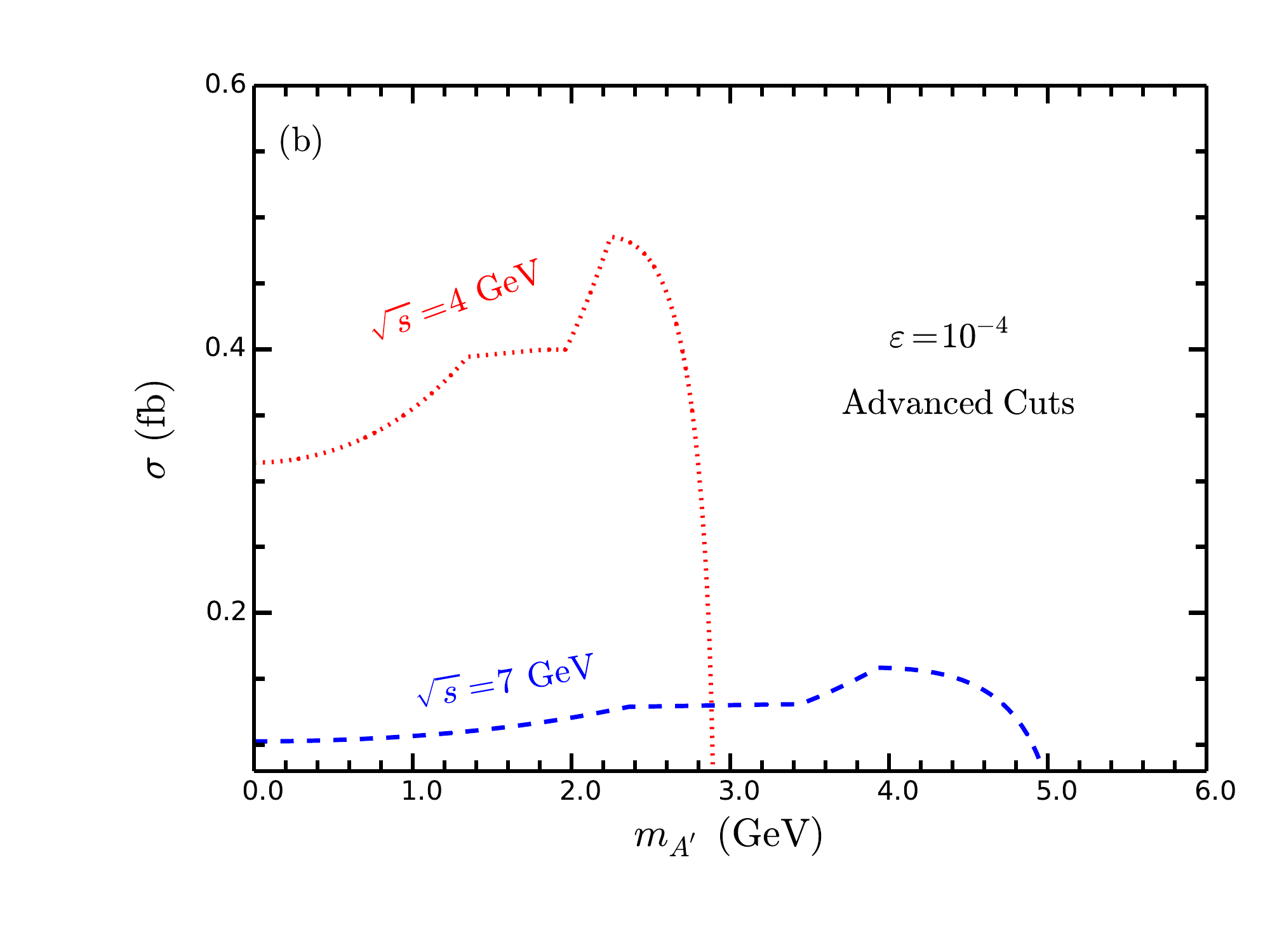}
\caption{Same as Fig. \ref{fig:xs-basic-cuts}. All the results are obtained by adopting the ``advanced cuts" and setting $\varepsilon=10^{-4}$.
}
		\label{fig:xs-adv-cuts} 
	\end{centering}
\end{figure}

In Fig. \ref{fig:xs-adv-cuts}  , we present 
the same results with Fig. \ref{fig:xs-basic-cuts} by using the ``advanced cuts" for the final photon. 
We can see that the production rates for dark photon and irreducible background in Fig. \ref{fig:xs-adv-cuts} (a)
have the same trend with Fig. \ref{fig:xs-basic-cuts} (a) when the colliding energy increases.
While the curves in Fig. \ref{fig:xs-adv-cuts} (b) have different shapes with   Fig. \ref{fig:xs-basic-cuts} (b).
As $m_{A^\prime}$ increases, the dark photon production rates starts to go up, reaches its maximum when
$m_{A^\prime}\simeq2.3\ (3.9)$ GeV for $\sqrt s$ = 4 (7) GeV, and then quickly goes down.

%
To simulate the detector effects on the final state particles, 
we smear the energy
 for the 
final state photon using Gaussian distributions 
which take into account the energy resolution of the EMC at 
the BESIII detector as \cite{Asner:2008nq} 
\be
\sigma(E)/E=2.3\%/\sqrt{E/{\rm GeV}}\oplus 1\%. 
\label{eq:deve}
\ee
For the EMC at STCF, we  assume the same energy resolution with BESIII.

\section{Results and calculation of the limit}
\label{sec:dplimit}
A large number of data have been accumulated by the BESIII 
detector at various running energies.
We summarize the BESIII data in Table \ref{tab:data}  since 2011 when
the monophoton trigger was implemented \cite{dywang}.
The data are arranged by the center-of-mass $\sqrt{s}$, of which the taking year and 
the luminosity are also listed \cite{besiii-data}. 
At each running energy, we simply estimate the upper bound on dark photon mixing strength  $\varepsilon$ at the 95\% confidence level (C.L.)
by solving the equation 
\be
{S_i^2(\varepsilon=\varepsilon^i_{95})  \over S_i(\varepsilon=\varepsilon^i_{95}) + B_i } = 2.71,
\ee 
where $S_i$ ($B_i$) is the number of signal (background) events and $\varepsilon^i_{95}$ is the 95\% C.L. upper bound on $\varepsilon$ at the $i$-th running energy. 
In order to  get the combined limit $\varepsilon_{95}$ by using all data 
listed in Table \ref{tab:data} at various energies at BESIII,
we solve the equation
\be
\sum_{i}{S_i^2(\varepsilon=\varepsilon_{95})  \over S_i(\varepsilon=\varepsilon_{95}) + B_i } = 2.71.
\ee

In the last column in Table \ref{tab:data}, we list $\varepsilon_{95}$ for the 
$m_{A^\prime}=1.5$ GeV  at each running energy at BESIII. 
The last row shows the limit combining all data between 2011 and 2018.
We can see that, when $m_{A^\prime}=1.5$ GeV, the upper bound on dark photon strength  $\varepsilon$ at the 95\% C.L.
can reach about $1.1\times 10^{-4}$ using the monophoton trigger data collected from 2011 to 2018.

\begin{table}[h] 
	\begin{center} 
		\begin{tabular}{|c|c|c|c|} 
			\hline 
			\hline
			Year& $\sqrt s$ (GeV)& ${\cal L}$ (fb$^{-1}$)   & $\varepsilon_{95}(\times 10^{-4})$\\
			\hline 
			2015 & 2.125 & 0.1       & 5.3  \\%
			2012+2018 & 3.097 & 1.8    & 1.7  \\%
			2017 & 3.515 & 0.46      & 3.6  \\%
			2011+2018 & 3.554 & 0.154   & 6.0 \\%
			2012+2018 & 3.686 & 1.0    & 2.7  \\%
			2011 & 3.773 & 1.99     & 2.1 \\%
			2017 & 3.872 & 0.22      & 5.6   \\%
			2011 & 4.009 & 0.5      & 4.0  \\%
			2016 & 4.18   & 3.1      & 2.1 \\%
			2013 & 4.23   & 1.05     & 3.1  \\%
			2013 & 4.26   & 0.83     & 3.5  \\%
			2017 & 4.28   & 3.8      & 2.0  \\%
			2012 & 4.36   & 0.5       & 4.4  \\%
			2014 & 4.42   & 1          & 3.4  \\%
			2014 & 4.6     & 0.5      & 4.7  \\%
			\hline 
			11-18 & -    & 17.004     & 1.1  \\%
			\hline\hline 
		\end{tabular} 
		\caption{The center-of-mass energy and corresponding luminosities collected from 2011 to 2018
			at the BESIII detector.
			The 95\% C.L.\ upper limits on $\varepsilon$ for the 
			$m_{A^\prime}=1.5$ GeV are listed in the last column. 
			The last row shows the result combining all data 
			between 2011 and 2018.}
		\label{tab:data} 
	\end{center} 
\end{table} 

\begin{figure}[htbp]
	\begin{centering}
		\includegraphics[width=0.7\columnwidth]{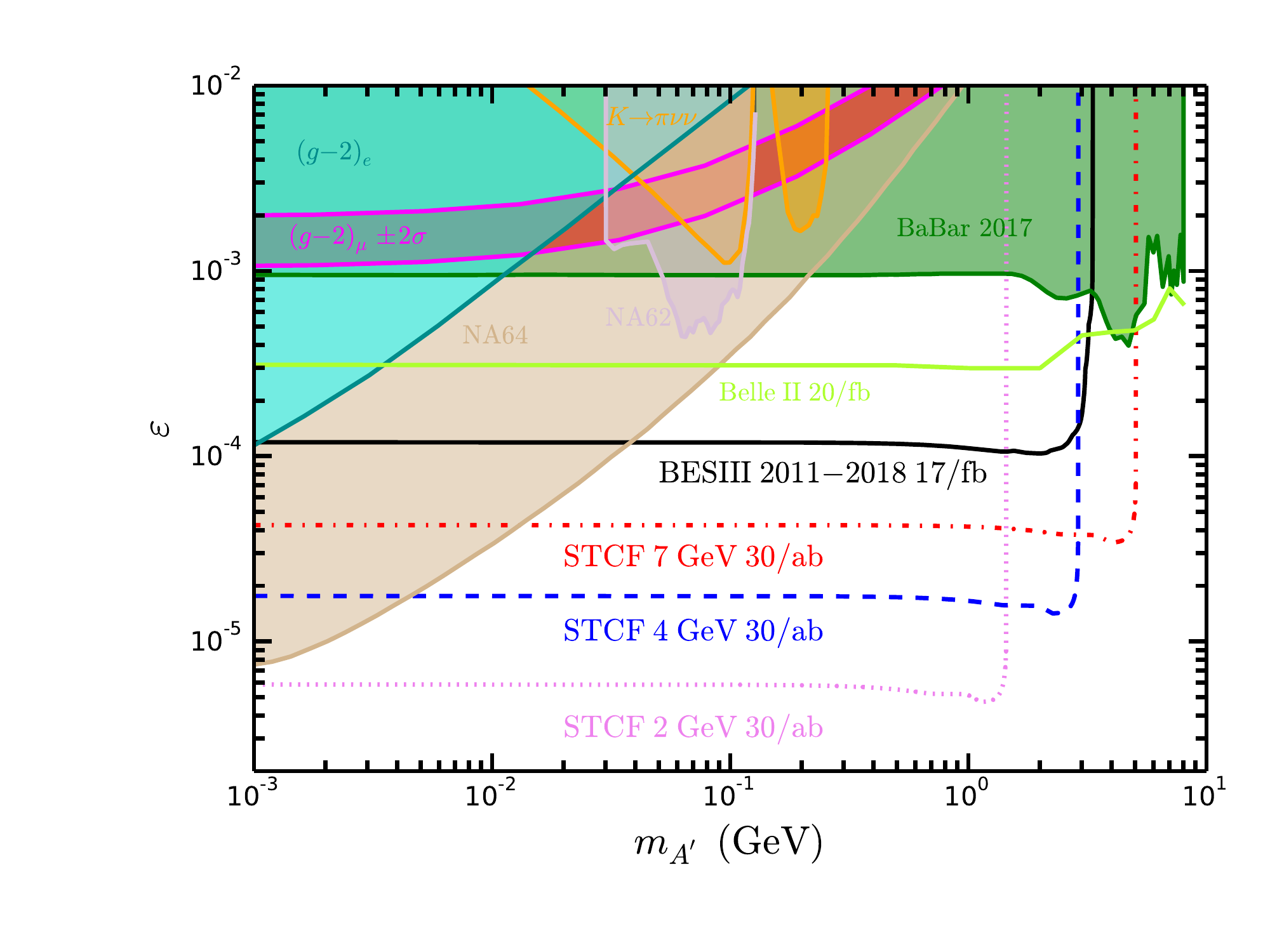}
		\caption{The expected 95\% C.L. exclusion limits on dark photon $A^\prime$ 
			mixing strength $\varepsilon$ as a function of the mass $m_{A^\prime}$
			at BESIII using the data collected during 2011-2018 (solid black).
			The STCF sensitivity curves are obtained assuming 30/ab integrated luminosity data 
			being collected at $\sqrt{s} =2$ GeV (dotted magenta), $\sqrt{s} =4$ GeV (dashed blue), $\sqrt{s} =7$ GeV (dot-dashed red), respectively. 
			The shaded regions show the existing bounds on $\varepsilon$ in channels where $A^\prime$
			is allowed to decay invisibly from the BaBar \cite{Lees:2017lec}, NA64 \cite{NA64:2019imj},
			NA62 \cite{CortinaGil:2019nuo} experiments, and the measurement for BR($K^+\to \pi^+\nu\bar{\nu}$) 
			by the E787 \cite{Adler:2001xv} and E949 \cite{Artamonov:2009sz} experiments,
			as well as the anomalous muon magnetic
			moment  $(g-2)\mu$ favored area \cite{Bennett:2006fi}.
			The projected upper limits on $\varepsilon$ for
			the process $e^+e^-\to \gamma A^\prime(\to {\rm invisible})$, for a 20 fb$^{-1}$ Belle II data set (solid green)
			\cite{Kou:2018nap} are also given.
		}
		\label{fig:eps95} 
	\end{centering}
\end{figure}

In Fig. \ref{fig:eps95}, we show the combined 95\% C.L. exclusion upper limits on $\varepsilon$ as a 
function of the mass $m_{A^\prime}$ via monophoton searches by using the data presented in Table \ref{tab:data} at BESIII
(solid black).
We also present the STCF sensitivity on $\varepsilon$ simply assuming about 30 ab$^{-1}$ data collected 
at $\sqrt{s} =2$ GeV (dotted magenta), $\sqrt{s} =4$ GeV (dashed blue), $\sqrt{s} =7$ GeV (dot-dashed red), respectively. 
The pre-existing  experimental constraints are also shown, which include 
the bounds in channels where $A^\prime$
is allowed to decay invisibly from the NA62 \cite{CortinaGil:2019nuo}, 
NA64 \cite{NA64:2019imj}, BaBar \cite{Lees:2017lec}, 
the measurement for BR($K^+\to \pi^+\nu\bar{\nu}$) 
by the E787 \cite{Adler:2001xv} and E949  \cite{Artamonov:2009sz} experiments,
as well as the anomalous muon magnetic
moment  $(g-2)\mu$ favored area \cite{Bennett:2006fi}. The projected upper limits on $\varepsilon$ for
the process $e^+e^-\to \gamma A^\prime(\to {\rm invisible})$, for a 20 fb$^{-1}$ Belle II data set (solid green)
\cite{Kou:2018nap} are also given.
We can see that BESIII  with about 17 fb$^{-1}$ data
can provide new leading upper limits to the mixing strength $\varepsilon$ of the dark photon 
in  the  mass  range $0.04~\text{GeV} \lesssim m_{A^\prime} \lesssim 3~\text{GeV}$,
of which the sensitivity is significantly better 
than future Belle II experiments with 20 fb$^{-1}$.
When $m_A^\prime=$ 1 GeV, the limits of $\varepsilon$ can be probed by BESIII down to $1.1\times 10^{-4}$,
which outperform the results from BaBar \cite{Lees:2017lec} about one order.
From the sensitivity on $\varepsilon$ at STCF with different collider energies, 
we can see that the low collider energy has better sensitivity than the high energy,
in spite of it touches smaller mass range.
For example, with 30 ab$^{-1}$ data, 2 GeV STCF can 
probe $\varepsilon$ down to $5.1\times 10^{-6}$ when $m_A^\prime=$ 1 GeV,
which outmatches 7 GeV STCF about 7 times.

\section{Constraints on light thermal dark matter}
\label{sec:dmlimit}

\begin{figure}[htbp]
	\begin{centering}
		\includegraphics[width=0.45\columnwidth]{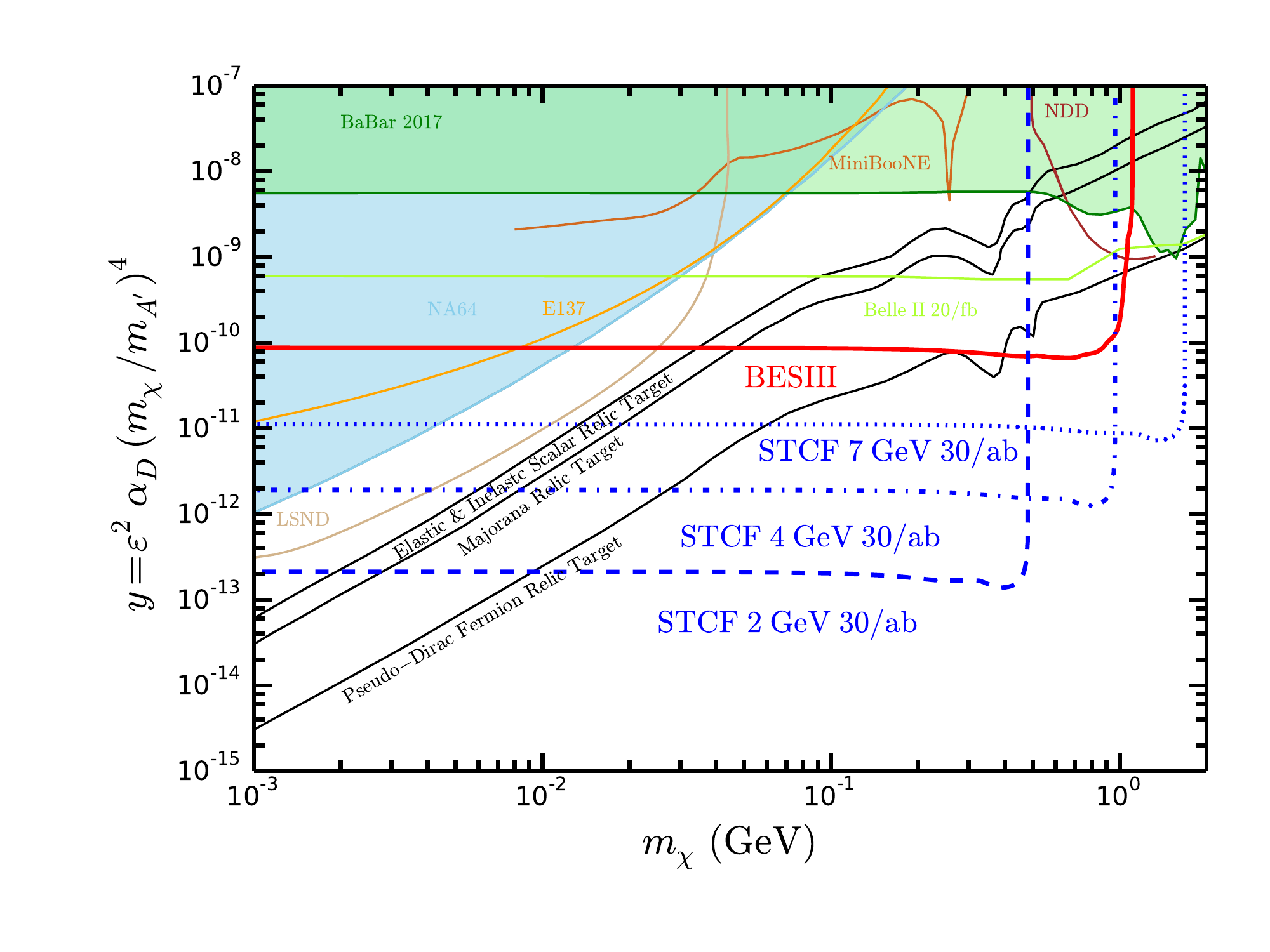}
		\includegraphics[width=0.45\columnwidth]{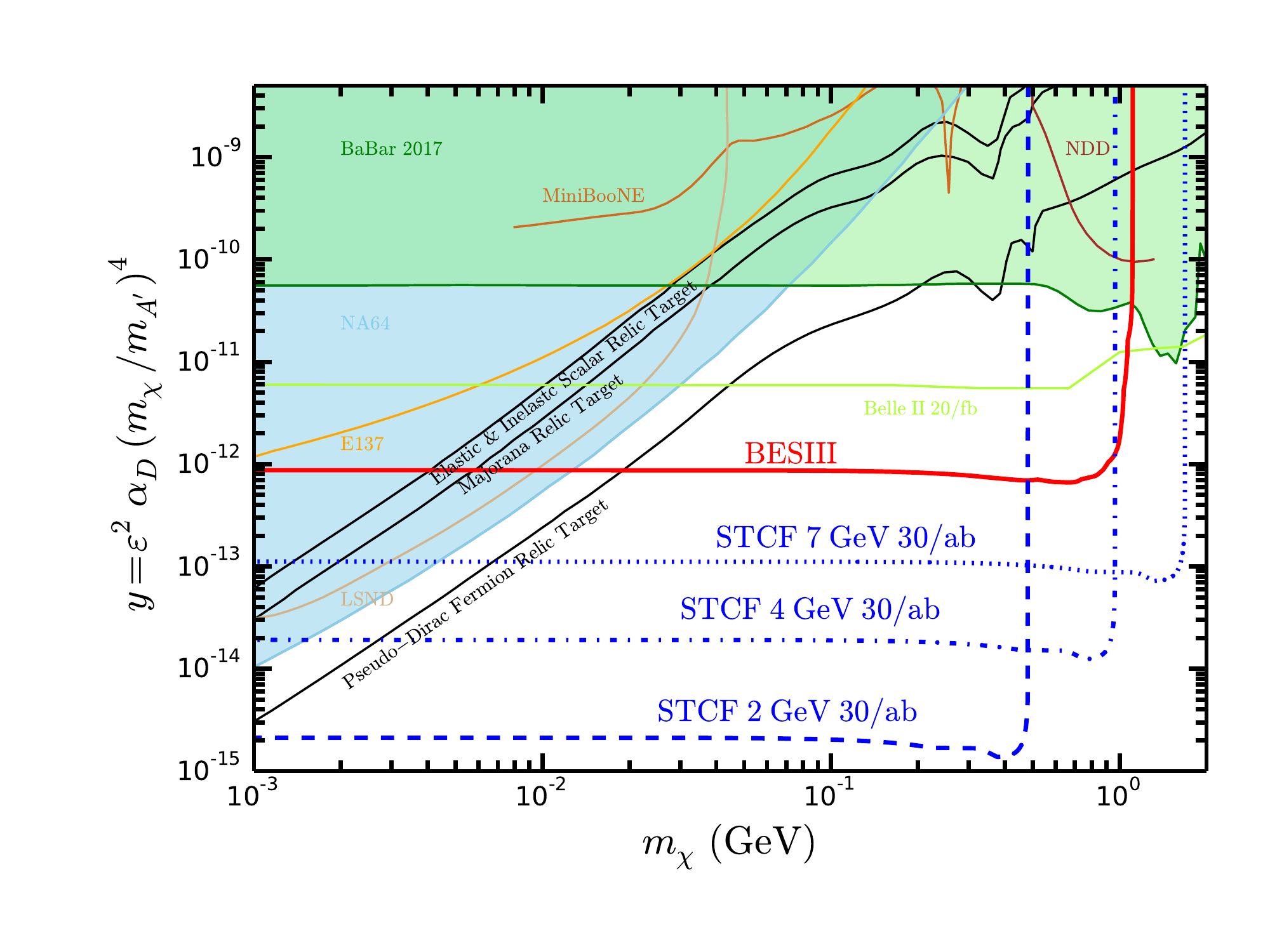}
		\caption{The expected 95\% C.L. exclusion limits on the parameter $y$ as 
		a function of the mass $m_{\chi}$
		from BESIII using the data collected during 2011-2018, as well as the future STCF.
		The STCF sensitivity curves are obtained assuming 30/ab integrated luminosity data 
		being collected at $\sqrt{s}=2$, 4, 7 GeV. The limits in the left panel are calculated 
		under the conventional assumption $m_{A^\prime}=3 m_{\chi}$ and $\alpha_D=0.5$. 
		In the right panel, the limits are shown for $\alpha_D=0.005$.
		The existing limits are obtained in Refs. \cite{Alexander:2016aln,NA64:2019imj, Izaguirre:2014bca,Izaguirre:2015yja,Izaguirre:2017bqb, Banerjee:2017hhz} from the 
		results of the NA64 \cite{NA64:2019imj}, LSND \cite{deNiverville:2011it,Batell:2009di}, E137 \cite{Batell:2014mga}, BaBar \cite{Lees:2017lec}, MiniBooNE \cite{Aguilar-Arevalo:2017mqx}, and nucleon direct detection (NDD) \cite{Essig:2012yx} experiments based on the missing mass, missing energy
		and missing momentum approaches. The favored parameters
		for the scalar, Majorana and pseudo-Dirac of LTDM to account for the observed relic DM density
		are shown as the solid lines \cite{Banerjee:2017hhz}.
			}
		\label{fig:y-chi} 
	\end{centering}
\end{figure}

In this section, we extend our discussions for the constraints on light thermal dark matter (LTDM).
The existence of thermal DM is arguably one of the most compelling possibilities, and 
has driven much of DM experiments over the past several years.
Among  the  thermal  DM  parameter space, 
the LTDM annihilating directly into SM particles  (the ``thermal relic target")
stick out for its predictiveness and testability \cite{Battaglieri:2017aum}. 
In the left panel of Fig. \ref{fig:y-chi}, we plot the expected 95\% C.L. values on 
the dimensionless DM annihilation cross section parameter $y=\varepsilon^2\alpha_D(m_\chi/m_{A^\prime})^4$
as function of the DM mass $m_{\chi}$ at BESIII and future STCF, 
where $\alpha_D=e_D^2/4\pi$,
under the conventional assumption $m_{A^\prime}=3 m_{\chi}$ and $\alpha_D=0.5$,
and compare them to different experimental exclusion regions.
The favored parameters for scalar, pseudo-Dirac (with a small splitting) and 
Majorana scenario of LTDM into account the observed relic DM density \cite{Banerjee:2017hhz} are also shown.
We can see that the direct search for the dark photon invisible decay  at BESIII via monophoton searches
excludes model of scalar, Majorana, and pseudo-Dirac (with a small splitting) DM for the 
mass region 0.03$\sim$1 GeV, 0.04$\sim$1 GeV and 0.4$\sim$1 GeV respectively.  
The choice of $\alpha_D=0.5$ is compatible with the bounds Ref. \cite{Davoudiasl:2015hxa} based on the running 
of the dark gauge coupling. However, it is important to note that the DM signal yields
in our analyses are primarily sensitive to $\varepsilon^2$, 
same as other accelerator experiments, such as NA64 \cite{Banerjee:2017hhz,NA64:2019imj}, 
different from $\varepsilon^4\alpha_D$ 
at the beam dump experiments, such as LSND \cite{deNiverville:2011it,Batell:2009di}, E137 \cite{Batell:2014mga},
MiniBooNE \cite{Aguilar-Arevalo:2017mqx}.
Therefore, our limits will be much stronger for sufficiently small values of $\alpha_D$.
In the right panel of Fig. \ref{fig:y-chi}, we present all the limits and bounds with 
 $m_{A^\prime}=3 m_{\chi}$ and $\alpha_D=0.005$.
 We can see that, for this or smaller values of $\alpha_D$, the model of scalar and Majorana DM 
 production via dark photon can be excluded by combining the NA64 \cite{NA64:2019imj}
 and  BaBar \cite{Lees:2017lec} limits.
 Furtherly combined with the limits from 2 GeV STCF with 30 ab$^{-1}$ data and BaBar \cite{Lees:2017lec},
 the model of pseudo-Dirac (with a small splitting) DM can also be excluded 
 for the entire plotted mass region.

\section{Summary}
\label{sec:sum}

In this work, we have proposed a search for invisible decay of dark photon  via  the  monophoton  signature  at
the  BESIII  detector and future STCF. The dark photon mixes with
the SM photon and decays dominantly invisible into light DM particles
$A^\prime \to \chi \bar{\chi}$.
New leading constraints on the 
mixing strength $\varepsilon$  can be obtained in the mass 
range $0.04~\text{GeV} \lesssim m_{A^\prime} \lesssim 3~\text{GeV}$ 
by using the current BESIII data. 
We also present  the sensitivity on  $\varepsilon$  at future STCF
with $\sqrt{s} =2, 4, 7$ GeV assuming about 30 ab$^{-1}$ data.
In addition, we discussed 
the constraints on light termal dark matter.
Using conventional choices, we provide the expected 95\% C.L. limits
on the dimensionless DM annihilation cross section
parameter $y$.
We find that the BESIII results can expand the search for
DM to $y$ values about two orders of magnitude smaller than
BaBar \cite{Lees:2017lec}.
For values $\alpha_D=0.005$ or smaller,
the model for scalar and Majorana DM production via dark photon
portal can be excluded by the combined results from direct searches 
of $A^\prime$ invisible decay in NA64 \cite{NA64:2019imj} and BaBar \cite{Lees:2017lec} experiments;
the model for pseudo-Dirac (with a small splitting)
can also be excluded by the combined results from 
2 GeV STCF with 30 ab$^{-1}$ data and BaBar \cite{Lees:2017lec}
for the mass region $0.001 \GeV \lesssim m_\chi \lesssim 1 \GeV$.


\acknowledgments
We thank Zuowei Liu for helpful discussions. 
This work was supported in part by the National Natural Science 
Foundation of China (Grants No. 11805001,
No. 11305001, No. 11575002, No. 11675033, No. 11747317， and No. 11875070)
and the Key Research Foundation of the Education
Ministry of Anhui Province of China (Grant No. KJ2017A032).

\end{document}